\documentclass[11pt,twoside]{article}
\usepackage{./asp2014}

\aspSuppressVolSlug
\resetcounters

\begin{document}

\title{Variability in the Milky Way: Contact binaries as diagnostic tools}
\author{Richard de Grijs,$^{1,2}$ Xiaodian Chen,$^{1,3}$ and Licai Deng$^3$
\affil{$^1$Kavli Institute for Astronomy \& Astrophysics and
  Department of Astronomy, Peking University, Yi He Yuan Lu 5, Hai
  Dian District, Beijing 100871, China; \email{grijs@pku.edu.cn}}
\affil{$^2$International Space Science Institute--Beijing, Nanertiao,
  Zhongguancun, Hai Dian District, Beijing 100190, China}
\affil{$^3$Key Laboratory for Optical Astronomy, National Astronomical
  Observatories, Chinese Academy of Sciences, 20A Datun Road, Chaoyang
  District, Beijing 100012, China}}

\paperauthor{Richard de Grijs}{}{0000-0002-7203-5996}{Peking University}{Kavli Institute for Astronomy and Astrophysics}{Beijing}{}{100871}{China}
\paperauthor{Xiaodian Chen}{}{}{Peking University}{Kavli Institute for Astronomy and Astrophysics}{Beijing}{}{100871}{China}
\paperauthor{Licai Deng}{}{}{National Astronomical Observatories, Chinese Academy of Sciences}{Key Laboratory for Optical Astronomy}{Beijing}{}{100012}{China}

\begin{abstract}
We used the 50 cm Binocular Network (50BiN) telescope at Delingha
Station (Qinghai Province) of Purple Mountain Observatory (Chinese
Academy of Sciences) to obtain simultaneous $V$- and $R$-band
observations of the old open cluster NGC 188. Our aim was a search for
populations of variable stars. We derived light-curve solutions for
six W Ursae Majoris (W UMa) eclipsing-binary systems and estimated
their orbital parameters. The resulting distance to the W UMas is
independent of the physical characteristics of the host cluster. We
next determined the current best period--luminosity relations for
contact binaries (CBs; scatter $\sigma < 0.10$ mag). We conclude that
CBs can be used as distance tracers with better than 5\%
uncertainty. We apply our new relations to the 102 CBs in the Large
Magellanic Cloud, which yields a distance modulus of
$(m-M_V)_0=18.41\pm0.20$ mag.
\end{abstract}

\section{Contact binary systems}

Contact binaries (CBs) are binary systems where both stellar
components overfill and transfer material through their Roche
lobes. They are rather common among the Milky Way's field stellar
population. In the solar neighborhood and the Galactic bulge, their
population density is approximately 0.2\%. In the Galactic disk it is
somewhat lower, on average, $\sim$0.1\% \citep{Rucinski06}. CBs can be
classified as early- and late-type systems; late-type CBs are also
known as W Ursae Majoris (W UMa) systems. Observational evidence
suggests that both binary components have very similar temperatures
although their masses differ, a conundrum known as Kuiper's paradox
\citep{Kuiper41}. As a solution to this paradox, \citet{Lucy68}
proposed convective common-envelope evolution as the key underlying
physical scenario of CB theory. Our modern view is that CBs have most
likely been formed through loss of angular momentum \citep{Stepien06,
  Yildiz13}.

\subsection{W Ursae Majoris systems}

The late-type, low-mass W UMa variables are, in essence, `overcontact'
binary systems. Both of their binary components usually rotate
rapidly, characterized by periods in the range from $P = 0.2$ days to
$P = 1.0$ day. One can indeed easily obtain complete, high-quality W
UMa light curves in just a few nights of observing time on relatively
small telescopes. In this contribution, we present such observations
of the six W UMa binary systems that reside in the old open cluster
(OC) NGC 188.

As highlighted above, W UMa systems are common in both OCs and the
Galactic field. This implies that they have great potential as
potential distance indicators. Indeed, approximately 0.1\% of the
F--K-type Galactic field dwarfs are late-type CBs \citep{Duerbeck84},
while in OCs their frequency may as high as $\sim 0.4$\%
\citep{Rucinski94}. If we could establish a reliable (orbital)
period--luminosity relation (PLR) for such W UMa systems, they might
potentially be employed to adopt a similarly important role as the
often used Cepheids or RR Lyrae variables in the context of measuring
the distances to old structures in the Milky Way. We note that while
distances to individual W UMa systems cannot be derived to the same
level of accuracy as those resulting from Cepheid analysis, the high
CB frequency in old stellar populations could potentially allow us to
overcome this disadvantage.

\subsection{NGC 188}

NGC 188 is located at a distance of $\sim 2$ kpc. It contains a
significant number of late-type CBs. Of these, seven W UMas near the
cluster's center were first found by \citet{Hoffmeister64} and
\citet{Kaluzny87}. Subsequently, \citet{Zhang02, Zhang04} surveyed
approximately 1 deg$^2$ around the center, yielding a CB haul of 16 W
UMa systems. \citet{Branly96} then used the Wilson--Devinney code to
calculate light-curve solutions for five of the central W UMas and
offered a discussion of the average W UMa distance compared with that
to the cluster as a whole. \citet{Liu11} and \citet{Zhu14} published
orbital solutions for EQ Cep, ER Cep, and V371 Cep, and for EP Cep, ES
Cep, and V369 Cep, respectively.

We observed NGC 188 over a continuous period of more than 2 months
using the 50 cm Binocular Network telescope \citep[50BiN;][]{Deng13}
at the Delingha Station (Qinghai Province, China) of Purple Mountain
Observatory (Chinese Academy of Sciences). We obtained simultaneous
time-series light-curve observations based on an unprecedented number
of 2700 frames, representing an effective observing time of 44 hr.
Details of the observations are included in \citet{chen16}. The
telescope's field of view, $20 \times 20$ arcmin$^2$, is adequate to
cover the cluster's central region.

To only select genuine cluster members, we performed detailed
radial-velocity and proper-motion analyses \citep{chen16}. We
eventually concluded that of our total sample of 914 stars, 532 stars
are probable cluster members. The latter delineate an obvious cluster
sequence in the color--magnitude diagram down to $V=18$ mag. We used
the Dartmouth stellar evolutionary isochrones \citep{Dotter08} to
ascertain the nature of the cluster members, adopting an age of 6 Gyr
and solar metallicity. We derived a distance modulus
$(m-M)_V^0=11.35\pm0.10$ mag and a reddening of
$E(V-R)=0.062\pm0.002$ mag.

\subsection{Distance determination}

\citet{Rucinski06} published a simple $M_V = M_V(\log P)$ calibration,
i.e. $M_V =(-1.5 \pm 0.8)-(12.0 \pm 2.0)\log P, \sigma =0.29$ mag,
based on his observations of 21 W UMa systems with good {\sl
  Hipparcos} parallaxes and All Sky Automated Survey (ASAS) $V$-band
photometry (maximum magnitudes). In \citet{chen16}, we established the
equivalent relationship using our own (50BiN) $V$-band data, combined
with the independently determined OC distance and the cluster's
average extinction.

We obtained accurate light-curve solutions for six W UMas of the NGC
188 variables. We used these to estimate the CBs' physical parameters,
including their mass ratios and the components' relative radii. We
subsequently estimated the distance modulus to the W UMa systems as a
whole, independently of the cluster distance. W UMas can be used to
derive distance moduli with an accuracy of often significantly better
than 0.2 mag. For this aspect of our distance-modulus analysis, we
excluded ER Cep given its low cluster-membership probability; in
addition, we suspect its nature as an eclipsing binary-type
system. For the remaining five W UMas---specifically, EP Cep, EQ Cep,
ES Cep, V369 Cep, and V370 Cep---we obtained a combined distance
modulus of $(m-M)_V^0=11.317 \pm 0.119$ mag. This value is indeed
comparable to the result from our isochrone fits,
$(m-M)_V^0=11.35\pm0.10$ mag, and also with previous results from the
literature. The accuracy resulting from our new analysis is much
better than that from application of the previously well-established
empirical parametric approximation.

To carefully check our results for the cluster as a whole and the
specific applicability of W UMas as distance tracer, we applied it to
the OC Berkeley 39. Based on four of the latter cluster's W UMas, we
derived a distance modulus of $(m-M)_V^0=13.09\pm0.23$ mag. This is
also in accordance with literature results. Thus, W UMas as potential
distance tracers have indeed significant advantages for the most
poorly studied clusters. Based on our initial analysis, we found that
five of our NGC 188 W UMa systems obey the overall W UMa PLR. Armed
with the latter, we were hopeful that W UMas could indeed play an
important role in measuring distances and to map Galactic structures
on more ambitious scales than done to date.

\section{Period--luminosity relations}

Although CBs are of order seven magnitudes fainter than the often used
Cepheid variables, within the same distance range their number is
three orders of magnitude larger. Cepheids trace young ($\la 20$
Myr-old) features; CBs are instead found in 0.5--10 Gyr-old stellar
populations. Although RR Lyrae stars are also members of structures
older than 1--2 Gyr, very few of the latter variables have been found
in either open clusters (OCs) or the solar neighborhood.

\begin{figure}
\centering
\includegraphics[width=100mm]{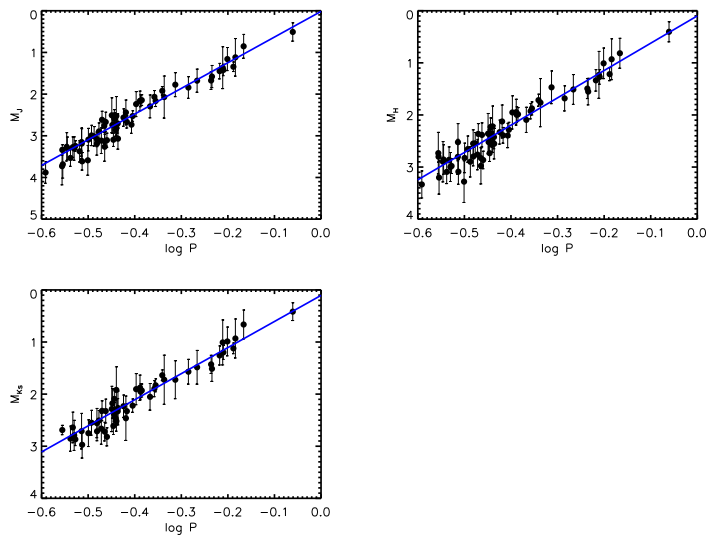}
\caption{Our new $JHK_{\rm s}$ CB PLRs (magnitudes as a function of
  period, in days), adapted from \citet{chen16b}.}\label{f4.fig}
\end{figure}

Since \citet{Eggen67}'s seminal work, these considerations and
observations have triggered a number of attempts at using CB
period--luminosity--color (PLC) relations to determine distances to
such old structures. \citet{Rucinski97} employed observations of 40 W
UMa-type CBs characterized by {\sl Hipparcos} parallaxes with an
accuracy in the corresponding distance moduli of $\epsilon_{M}<0.5$
mag to improve their PLC relation. Subsequently, \citet{Rucinski06}
derived a luminosity function composed of CBs sourced from the ASAS
data. He then explored the viability of a $V$-band PLR. However, his
`PLR' exhibited only a weak correlation and was affected by large
uncertainties and significant scatter.

We collected CBs in OCs and CBs with accurate {\sl Hipparcos}
parallaxes \citep{chen16b}, aiming at establishing more reliable
PLRs. Our full sample contains 6090 CBs from the General Catalog of
Variable Stars and the ASAS. Our corresponding OC sample contains 2167
OCs. We made a special effort to exclude foreground and background
CBs, requiring that (i) any suitable CB must be located inside the
core radius of its host OC; (ii) the CB's proper motion must be
located within the $2\sigma$ distribution of that of its host OC; and
(iii) the CB's age must be similar to that of its host OC,
i.e. $\Delta \log (t\mbox{ yr}^{-1}) <0.3$. Our final sample selection
consisted of 42 high-probability OC CBs. Combined with four nearby
moving-group CBs and 20 W UMa-type CBs with accurate {\sl Hipparcos}
parallaxes, we hence used a sample of 66 CBs to determine the
$JHK_{\rm s}$ PLRs, i.e.,

\begin{displaymath}\label{equation3}
  \begin{aligned}
   M_{J_{\rm max}}^{\rm late} &= (-6.15 \pm 0.13) \log P+(-0.03 \pm 0.05), \sigma_J=0.09,(\log P < -0.25); \\
   M_{J_{\rm max}}^{\rm early} &= (-5.04 \pm 0.13) \log P+(0.29 \pm 0.05), \sigma_J=0.09,(\log P > -0.25); \\
   M_{H_{\rm max}} &= (-5.22 \pm 0.12) \log P+(0.12 \pm 0.05), \sigma_H=0.08; \\
   M_{K_{\rm s,max}} &= (-4.98 \pm 0.12) \log P+(0.13 \pm 0.04), \sigma_K=0.08.
   \end{aligned}
\end{displaymath}

These PLRs result in distances that are as accurate as those derived
from the $JHK_{\rm s}$ Cepheid PLRs (scatter $\sigma < 0.10$ mag): see
Fig. \ref{f4.fig}. In fact, these relations are the first PLRs for
early-type CBs thus far established at near-infrared wavelengths. in
order to verify the accuracy of our PLRs, we carefully investigated
the CBs' period--color relations. The latter can be employed to get
rid of unreliable CBs. Near-infrared PLRs are more accurate and
significantly less sensitive to extinction and metallicity variations
than $V$-band PLRs.

Combining the $JHK_{\rm s}$ PLRs, we derived the distances to our
sample of 6090 CBs. The resulting accuracy is high: 90\% of our sample
CBs have distance errors of less than 5\%, and 95\% have distance
uncertainties of less than 10\%. The remaining 5\% may be CBs
associated with poor-quality photometry, variables affected by high or
complicated differential extinction, or objects that could have been
misidentified as CBs, e.g. semi-detached binaries and---for small
amplitudes---RR Lyrae and ellipsoidal binaries.

\subsection{Application to the Large Magellanic Cloud}

\citet{Graczyk11} published a catalog of 26,121 eclipsing binaries in
the Large Magellanic Cloud (LMC), which ad been identified based on
visual inspection of the Optical Gravitational Lensing Experiment III
catalog. Their 1048 type-EC eclipsing binaries are CBs, although they
only included CBs with long periods ($\log P >-0.2$ [days]). To select
CBs that can be used as reliable distance tracers, we adopted our
period--color selection and imposed period limits of $-0.13<\log P
<0.2$. Here, the upper limit is at the long-period end of the CB
distribution and the lower limit is the magnitude limit used for
detecting LMC CBs.

This resulted in a total sample of 102 LMC CBs and a distance modulus
of $(m-M_{V})_0=18.41\pm 0.20$ mag. This is first distance to the LMC
based on CBs. It is entirely consistent with the current best LMC
distance modulus \citep{de Grijs14}, $(m-M)_0=18.49 \pm 0.09$
mag. 


\end{document}